\documentclass[%
reprint,
superscriptaddress,
groupedaddress,
pra,
]{revtex4-1}

\usepackage[utf8]{inputenc} % allow utf-8 input
\usepackage[T1]{fontenc}    % use 8-bit T1 fonts
\usepackage{hyperref}       % hyperlinks
\usepackage{url}            % simple URL typesetting
\usepackage{booktabs}       % professional-quality tables
\usepackage{amsfonts}       % blackboard math symbols
\usepackage{nicefrac}       % compact symbols for 1/2, etc.
\usepackage{microtype}      % microtypography
\usepackage{lipsum}		% Can be removed after putting your text content
\usepackage{graphicx}
\usepackage{svg}
\usepackage{natbib}
\usepackage{amsmath}
\usepackage{mathrsfs}

\begin{document}

%\preprint{APS/123-QED}

\title{Scaling law from orbital angular momentum conservation in harmonic and high-order harmonic generation driven by spatiotemporal light fields}

\author{Miguel A. Porras}
\affiliation{Complex Systems Group, ETSIME, Universidad Politécnica de Madrid, Ríos Rosas 21, 28003 Madrid, Spain}
\email{miguelangel.porras@upm.es}

\author{Marcos G. Barriopedro}
\affiliation{Instituto de Fusión Nuclear ``Gillermo Velarde'', Universidad Politécnica de Madrid (ETSII),
C. de José Gutiérrez Abascal 2, 28006 Madrid, Spain}
\altaffiliation{Also at Complex Systems Group}

\author{Rodrigo Mart\'{\i}n-Hern\'andez}
\affiliation{Grupo de Investigación en Aplicaciones del Láser y Fotónica, Departamento de Física Aplicada,
Universidad de Salamanca, 37008, Salamanca, Spain}
\affiliation{Unidad de Excelencia en Luz y Materia Estructuradas (LUMES), Universidad de Salamanca, Salamanca, Spain}

\begin{abstract}
Nonlinear photon upconversion processes driven by diverse forms of structured light are receiving increasing attention. In harmonic and high-order harmonic  generation (HG and HHG) with Laguerre-Gauss (LG) beams, linear scaling the driver topological charge (TC) with the harmonic order is equivalent to driver orbital angular momentum (OAM) per photon scaling, and constitutes a proof of OAM conservation. However, with generic driving fields, such as non-LG vortices or spatiotemporal optical vortices, TC and OAM per photon may scale or not in a process in which the OAM is conserved. We find the physical magnitude that scales with generality when the OAM, either  longitudinal or transverse, or its intrinsic part, is conserved. This new rule allows for the wealth of phenomena observed in HHG that are unintelligible from the rigid LG rule.
\end{abstract}

%Nonlinear photon upconversion processes, perturbative or non-perturbative, driven by diverse forms structured light are receiving increasing attention. In harmonic generation with Laguerre-Gauss beams  scaling of the topological charge of the driver with the harmonic order is equivalent to scaling of the orbital angular momentum (OAM) per photon, and a test of OAM conservation, but . With generic spatiotemporal fields, TC and OAM (or intrinsic OAM), scaling or nonscaling, are unrelated to OAM conservation. We find the physical magnitude that scales with harmonic order with generality as a result of OAM conservation. This study sets a basis for a correct understanding of present and future analysis and experiments of nonlinear processes with structured light carrying OAM.

\maketitle

\section{Introduction}

Shortly after the discovery that Laguerre-Gauss (LG) beams possess orbital angular momentum (OAM) directed along the propagation direction  \cite{allen92}, interest arose in exploring its role in nonlinear phenomena. It is now well-known that LG beams transmit OAM to the nonlinearly generated second-order \cite{dholakia1996,padgett1997}, third-order \cite{olivier08}, and high-order harmonics  \cite{gariepy2014,geneaux2016,gauthier2017,hernandezgarcia2013}. LG beams also posses a phase singularity, or vortex. This is a topological feature characterized by a winding number, or topological charge (TC), measuring the number of times the phase changes $2\pi$ in one turn about the singularity. For LG beams, and more generally, vortex beams with cylindrical symmetry, OAM per photon and TC $\ell$ are related by $\hbar \ell$.

A fundamental question is the conservation of the OAM in nonlinear phenomena. In the generation of a harmonic of order $q$ of frequency $\omega_q=q\omega_1$ from a fundamental frequency $\omega_1$, $\hbar \omega_q =q\hbar \omega_1$ expresses the conservation of the energy of the $q$ photons that are upconverted to a single photon. Conservation of the OAM requires $\hbar \ell_q= q\hbar \ell_1$ in the upconversion ---the OAM per photon scales with harmonic order---, which is equivalent to $\ell_q=q\ell_1$ ---the TC scales with harmonic order. Thus, counting TC scaling with harmonic order is commonly regarded a proof of OAM conservation with LG beams \cite{padgett1997,dholakia1996,gariepy2014,gariepythesis}.

Recently, research has expanded to nonlinear processes driven by spatiotemporal structured fields \cite{2023_bliokh_Roadmap,2023_shen_Roadmap} with topological features, for example, to second-order \cite{gui2021,hancock2021} and third-order \cite{wang23third} harmonic generation (HG), and more recently to high-order harmonic (HHG) \cite{fang2021,Martin-Hernandez2025spatiotemporal,porras2025arxivPRL} driven by spatiotemporal optical vortices (STOVs). As LG beams, STOVs carry an OAM, but directed transversally to the propagation direction, and a spatiotemporal phase singularity, also characterized by a TC. In \cite{gui2021,hancock2021}, the TC of the second-order harmonic doubles that of the driver STOV, which is considered a verification of OAM conservation \cite{hancock2021}. In  \cite{fang2021,Martin-Hernandez2025spatiotemporal}, the TC also scales with the high-order harmonic order, but in the more recent study of HHG driven by focused STOVs  \cite{porras2025arxivPRL}, it does not. The TC is the same as that of the driver for all the harmonics, at the same time that the OAM is conserved.

It is today understood that TC is generally unrelated to OAM \cite{berry2022}. The former quantifies a topological property of light and the latter a mechanical property \cite{berry2022}. A variety of extreme situations with vanishing TC and non zero longitudinal OAM, and vice versa, are reported in \cite{berry2022}. STOVs may reverse the sign of their TC in free space propagation while the OAM remains constant \cite{porras2023transverse}, or can have arbitrary TC and zero OAM \cite{porras2023propagation}.
On the other hand, while the OAM per photon in LG beams is a well-defined amount of OAM carried by each photon \cite{zeilinger01}, in general it is an average quantity evaluated from a classical field, which can change in nonlinear processes. As seen here, harmonic OAM per photon generally differs from $q$ times that of the driver in a process where the OAM is conserved.

In this paper, we find the physical magnitude that scales with the harmonic order in HG or HHG processes driven by arbitrary spatiotemporal fields as a result of OAM and energy conservation. It is the ratio of converted OAM per converted photon number that is always $q$ times that of the input driver. This scaling law holds for the longitudinal OAM (l-OAM), the transverse OAM (t-OAM), and its intrinsic part, in case that they are conserved. This general scaling law reduces to the more intuitive scaling law of the OAM per photon when the latter does not change in the process, which in turn reduces to the TC scaling law if OAM per photon and TC are proportional.

We first derive the scaling law in HG and illustrate its validity with a couple of examples where the l-OAM and t-OAM are conserved. We then show that the same scaling law expresses OAM conservation  in HHG. We are then able to demonstrate from an analytical model l-OAM and t-OAM conservation in HHG in a thin gas-jet driven by arbitrary spatiotemporal fields. We corroborate conservation using advanced numerical simulations based on the macroscopic strong field approximation in examples where the short-range LG scaling rule would indicate otherwise.

\section{Scaling law in HG}

Assume that the relevant component of the OAM, $L$, is conserved in a nonlinear process of HG along with the energy $W$. The driving and harmonic fields are assumed to be quasimonochromatic of carrier frequencies $\omega_1$ and $\omega_q =q\omega_1$. The energy at the beginning and the end of the process are related by $W_1^{\rm in}+W_q^{\rm in} =W_1^{\rm out}+W_q^{\rm out}$.
Since each photon carries, on average, an energy $\hbar\omega_i$, the average number of photons is $N_i=W_i/\omega_i$ (with units of $\hbar$), and the conservation of energy reads  $N_1^{\rm in}+qN_q^{\rm in} =N_1^{\rm out}+qN_q^{\rm out}$, similar to a Manley-Rowe relation, or $N_1^{\rm in}-N_1^{\rm out}=q(N_q^{\rm out}-N_q^{\rm in})$. Analogously, from OAM conservation, $L_1^{\rm in}+L_q^{\rm in} =L_1^{\rm out}+L_q^{\rm out}$, or $L_1^{\rm in}-L_1^{\rm out}=L_q^{\rm out}-L_q^{\rm in}$. We then obtain
\begin{equation}\label{scaling1}
    \frac{L_q^{\rm out}-L_q^{\rm in}}{N_q^{\rm out}-N_q^{\rm in}}=q  \frac{L_1^{\rm out}-L_1^{\rm in}}{N_1^{\rm out}-N_1^{\rm in}}.
\end{equation}
The fractions in the two members have the same meaning, involving only the harmonic field in the first member, and the driver in the second. Therefore, Eq. (\ref{scaling1}) is the scaling law that is satisfied with generality if the OAM is conserved along with the energy. It is the ratio of converted OAM to converted photon number that scales with the harmonic order in the process. If the harmonic field is not seeded, Eq. (\ref{scaling1}) becomes
\begin{equation}\label{scaling2}
    \frac{L_q^{\rm out}}{N_q^{\rm out}}=q  \frac{L_1^{\rm out}-L_1^{\rm in}}{N_1^{\rm out}-N_1^{\rm in}} \equiv q\frac{\Delta L_1}{\Delta N_1}.
\end{equation}
The ratio of converted OAM to converted photon number of the harmonic coincides with the OAM per photon in this case.
It is evident from Eq. (\ref{scaling2}) that the OAM per photon of the harmonic is not equal to $q$ times the OAM per photon of the input driver, $L_1^{\rm in}/N_1^{\rm in}$, but to the ratio of converted OAM $\Delta L_1$ to converted photon number $\Delta N_1$ of the driver. The exception with arbitrary fields is that the driver is completely depleted ($L_1^{\rm out}=0$, $N_1^{\rm out}=0$), which rarely occurs.

\begin{figure*}
    \centering
    \includegraphics[width=\linewidth]{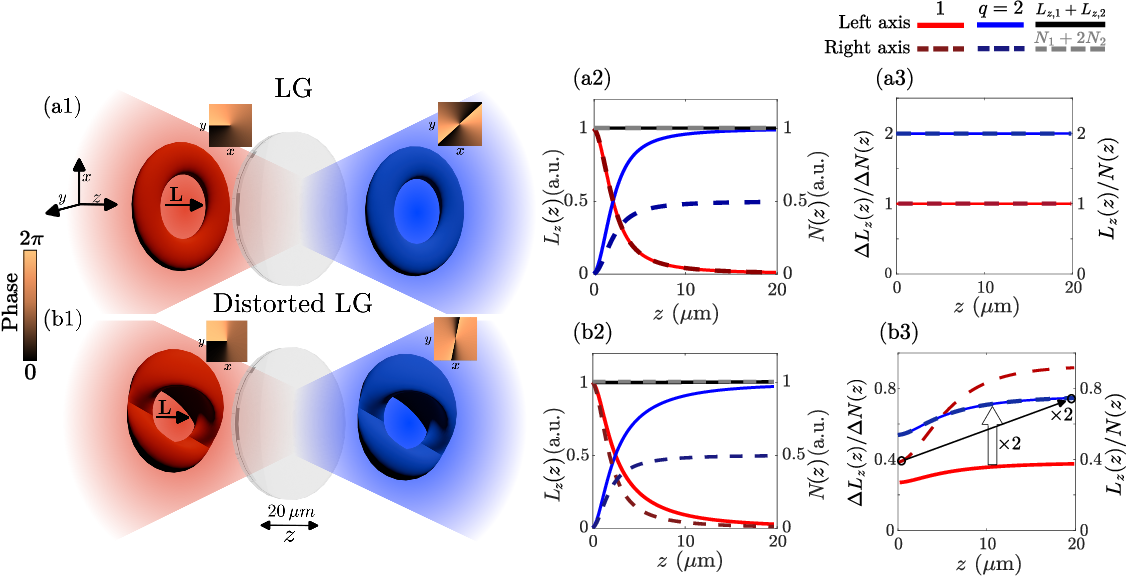}
    \caption{(a1) Left: Iso-intensity surface and phase of fundamental pulsed LG $\psi_1(x,y,t',0) = A(\frac{x}{x_0} + i\eta \frac{y}{y_0})e^{-\frac{x^2}{x_0^2}-\frac{y^2}{y_0^2}-\frac{t'^2}{t_0^2}}$ (left) at $\omega_1=2.35$ rad/fs, with $\ell=1$, $x_0 = 0.5$ mm, $y_0 = 0.5$ mm, $t_0 = 200$ fs, $\eta= 1$ and $A$ adjusted to obtain $W = 6.5$ mJ. Right: Iso-intensity surface and phase of the generated second harmonic (right) at $\omega_2 = 2\omega_1 = 4.7$ rad/fs at $z=20$ $\mu$m. The effective nonlinear coefficient is $\kappa = 1.466\times10^{-4}$ V$^{-1}$, and $n_1 = n_2 = 1.661$ is assumed. (b1) The same but with fundamental distorted LG by setting the distotion parameter $\eta= 0.2$, and $A$ adjusted to obtain $W = 6.5$ mJ. (a2, b2) l-OAM (solid) and number of photons (dashed) for the fundamental and the harmonic fields as functions of the medium thickness, for the LG and distorted LG. (a3, b3) ratio of converted l-OAM to converted photon number (solid) and l-OAM per photon (dashed) for the fundamental (red) and the harmonic (blue) as functions of the medium thickness, for the LG and distorted LG.}
    \label{fig:FIG1}
\end{figure*}

It follows from Eq. (\ref{scaling2}) that the more intuitive OAM per photon scaling, $L_q^{\rm out}/N_q^{\rm out} = q L_1^{\rm in}/N_1^{\rm in}$, only holds without complete depletion if $L_1^{\rm out}/N_1^{\rm out} = L_1^{\rm in}/N_1^{\rm in}$, i.e., if the OAM per photon of the driver does not change in the process. This is the case with longitudinal vortex fields with cylindrical symmetry about the propagation axis, e.g., LG beams, in a nonlinear process that does not break this symmetry, or other fields with particular symmetries.  In such cases, the general scaling law in Eq. (\ref{scaling2}) simplifies to  $L_q^{\rm out}/N_q^{\rm out} = q L_1^{\rm in}/N_1^{\rm in}$. For a LG beam of TC $\ell$, for instance, $L/N= \ell$ (in units of $\hbar$), and the OAM per photon scaling is also equivalent to the TC scaling $\ell_q=q\ell_1$.
This fits the quantum picture in which $q$ photons carrying each a well-defined value of the OAM $\ell_1$ are absorbed and a single photon with well-defined OAM $q\ell$ is emitted.

However, for a generic spatiotemporal driver, photons do not carry a well-defined OAM. The OAM per photon is an average property inferred from a classical field, a property that generally changes in the upconversion process. In this general situation, $L_q^{\rm out}/N_q^{\rm out} \neq q L_1^{\rm in}/N_1^{\rm in}$, and it is the scaling law in Eq. (\ref{scaling2}) that is satisfied if the OAM is conserved.

\subsection*{Example with l-OAM}

Figure \ref{fig:FIG1} illustrates the above facts in phase matched second-order ($q=2$) harmonic generation (SHG) driven by (a1) a pulsed LG beam of TC $\ell_1=1$ and (b1) a generic longitudinal vortex beam of the same TC but without cylindrical symmetry (obtained by deforming a LG beam) in a transparent nonlinear crystal of different thicknesses $z$, so that "in" is now $z=0$, "out" is $z$, $\Delta L_1(z) =L_{z,1}(z)- L_{z,1}(0)$, and $\Delta N_1(z)=N_1(z)-N_1(0)$. The drivers and medium parameters (see caption) are chosen such that they are almost completely depleted at the maximum thickness $z$, at the same time that diffraction effects remain negligible. Under these conditions, SHG is ruled by the coupled mode equations $\partial \psi_1/\partial z =i\kappa \psi_1^\star\psi_2$ and $\partial \psi_2/\partial z = i\kappa\psi_1^2$, which we have solved numerically. The second of the two coupled mode equations imposes that the second harmonic emerges with TC $\ell_2=2$, as seen in (a1) and (b1). The l-OAM of the fundamental and second harmonic fields are calculated from
\begin{equation}\label{eq:Lz}
L_{z,j} = \frac{\varepsilon_i}{2k_j}{}\int\psi_j^* [-i(x\partial_y-y\partial_x)]\psi_j dxdy dt',
\end{equation}
where, $(x,y)$ are Cartesian coordinates in transversal planes, $t'=t-z/v_j$ is the local time, $\varepsilon_j= \varepsilon_0 n_j^2$, $k_j= \omega_j n_j/c$ the propagation constants, $v_j=\omega_j/k_j$, and we choose $n_j=n$, $j=1,2$. Since the pulse energy is
\begin{equation}\label{eq:energy}
W_j=\frac{1}{2}\varepsilon_0 n_j c \int |\psi_j|^2 dxdy dt',
\end{equation}
and $N_j=W_j/\omega_j$, the l-OAM per photon is
\begin{equation}\label{eq:Lzphoton}
\frac{L_{z,j}}{N_j}= \frac{\int\psi_j^* \hat L \psi_j dxdy dt'}{\int |\psi_j|^2dxdy dt'} ,
\end{equation}
where we write $\hat L=-i\left(x\partial_y-y\partial_x\right)$ for convenience,
and a similar expression for $\Delta L_{z,j}/\Delta N_{z,j}$ with the difference of the integrals at $z$ and $0$ in the numerator and denominator.

The total l-OAM $L_z =L_{z,1}+ L_{z,2}$ [black lines in Figs. \ref{fig:FIG1} (a2) and (b2)] is conserved in both cases, with decreasing $L_{z,1}$ [red curves in (a2) and (b2)] and increasing $L_{z,2}$ [blue curves in (a2) and] with increasing medium thickness $z$. The number of photons $N_1$ decreases with $z$ [dashed dark red curves in (a2) and (b2)], and $N_2$ increases by up to half in both cases [dashed dark blue curves in (a2) and (b2)] with conserved $N_1+2N_2$ [dashed gray lines in (a2) and (b2)], i.e., conserved energy.

With the LG pulse, the scaling law from l-OAM conservation $L_{z,2}/N_2(z) = 2 \Delta L_{z,1}(z)/\Delta N_{1}(z)$ [blue and red solid lines in (a3)] is equivalent to $L_{z,2}(z)/N_2(z) = 2 L_{z,1}(0)/N_{1} (0)$ [blue dashed line and red dashed line at $z=0$ in (a3)], and equivalent to  $\ell_2=2\ell_1$, since $L_{z,1}/N_1$ does not change in the process and is proportional to $\ell_1$.

With the generic longitudinal vortex beam, $L_{z,2}(z)/N_2(z) = 2\Delta L_{z,1}(z)/\Delta N_{1}(z)$ in a medium of any thickness $z$ as a result of l-OAM conservation [blue and red solid curves in (a3)], as indicated by the vertical arrow, presenting different values for each output $z$. For no value of $z$ the l-OAM per photon $L_{z,2}(z)/N_2(z)$ of the output harmonic is twice the l-OAM per photon $L_{z,1}(0)/N_1(0)$ of the input driver [blue dashed curve and red dashed curve at $z=0$ in (b3)], except if the driver is depleted, as indicated by the diagonal arrow. The l-OAM per photon of both the driver and harmonic fields change as the SHG process inhomogeneously deforms them. It may be, as in (b3), that the l-OAM per photon of the output harmonic field is smaller than that of the output driver for some $z$, even if its TC is double.
$$\star\star\star$$
The scaling law in Eq. (\ref{scaling2}) holds for any component of the OAM that is conserved along with the energy. If the propagation direction $z$ is the direction of momentum $\bf P$, the OAM along this direction, the l-OAM, is purely intrinsic, meaning that it is independent of the particular parallel $z$ axis \cite{ONeil}. In contrast, the t-OAM is not purely intrinsic but it depends on the particular transverse axis. Yet, one can define a t-OAM that is independent of the particular parallel transverse axis, or intrinsic t-OAM, by evaluating the t-OAM about a transverse axis passing through a physical definition of the wave center. Two definitions have been proposed: the center of the energy density \cite{porras2023transverse} and the center of the photon wave function \cite{Bliokh2023}. With any choice, the intrinsic t-OAM may be conserved or not in a nonlinear process. In the case of conservation, the scaling law in Eq. (\ref{scaling2}) also holds for them.

\begin{figure*}
    \centering
\includegraphics[width=\linewidth]{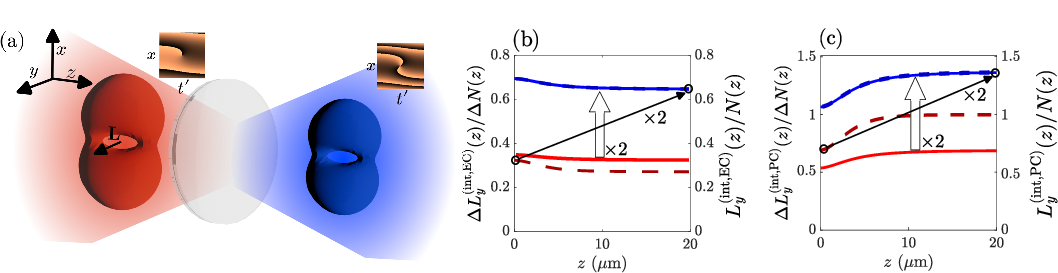}
    \caption{(a) Left: Iso-intensity surface and phase of the non-elliptical STOV $\psi_1(x,y,t',0) \propto(t'/\eta t_0 +ix/x_{0,f})\exp(t'^2/t_0^2-x^2/x_{0,f}^2-y^2/y_{0,f}^2)\exp[ik_1(x_0^2+y_0^2)/2f]$ at $\omega_1=2.35$ rad/fs, of charge $\ell=1$ and distortion parameter $\eta= 0.4$ at the entrance $z=0$ of the nonlinear medium, obtained by focusing with focal length $f=250$ mm the tilted lobed field  $A (t/\eta t_0 + x/x_0)\exp(-x^2/x_0^2-y^2/y_0^2-t'^2/t_0^2)$ with $x_0 = y_0 = 1.0$ mm, $t_0 = 200$ fs, and $A$ adjusted to obtain $W = 6.5$ mJ. The focused STOV width is $x_{0,f}=y_{0,f} =2f/k_1x_0$. Right: Iso-intensity surface and phase of the generated second harmonic at $\omega_2 = 2\omega_1 = 4.7$ rad/fs at $z=20$ $\mu$m. The nonlinear coefficient is $\kappa = 1.18\times10^{-5}$ V$^{-1}$ and $n_1 = n_2 = 1.661$. (b, c) Ratio of converted intrinsic t-OAM per converted photon number (solid) and intrinsic t-OAM per photon (dashed) for the fundamental (red) and the second harmonic (blue) as functions of the medium thickness, for the respective choices of the EC and the PC to extract the intrinsic t-OAM.
%$\gamma=0.94234$
    }
    \label{fig:FIG2}
\end{figure*}

\subsection*{Example with t-OAM}

Consider now SHG under the same conditions as in the example of Fig. \ref{fig:FIG1}, i.e., phase matched SHG and negligible diffraction effects up to the maximum medium thickness $z$ for almost complete depletion (then ruled by the same coupled mode equations as above), but the fundamental field at the entrance plane $z=0$ is a distorted, non-elliptical STOV of TC $\ell_1=1$ with t-OAM along the $y$ direction, formed at the focal plane of a lens illuminated by a tilted two-lobe field (see caption for their expressions). The distorted STOV and a particular generated second harmonic field, also a distorted  STOV of TC $\ell_2=2$, are shown in Fig. \ref{fig:FIG2}(a). Since the propagation direction and momentum of these fields are parallel ($P_x=P_y=0$), the t-OAM with respect to the $y$ axis passing through $x=0,z=0$ can be evaluated from the simplified formula \cite{porras2024clarification,porras2025arxivPRL}
\begin{equation}\label{eq:Ly}
L_{y,j} = -\frac{\varepsilon_i}{2}\int\psi_j^* x\psi_j dxdy dt',
\end{equation}
the intrinsic t-OAM with respect to the energy centroid (EC) from \cite{porras2024clarification}
\begin{equation}\label{eq:LyintEC}
L_{y,j}^{(\rm int,EC)} =\frac{\varepsilon_j}{2k_j}\int\psi_j^* i v_jt'\partial_x\psi_j dxdy dt',
\end{equation}
and with respect to the photon centroid (PC) from \cite{porras2025arxivPRL}
\begin{equation}\label{eq:LyintPC}
L_{y,j}^{(\rm int,PC)} =\frac{\varepsilon_j}{2k_j}\int\psi_j^* i( v_jt'-x\partial_{v_j t'})\partial_x\psi_j dxdy dt'.
\end{equation}
As with the l-AOM, the t-OAM and the intrinsic t-OAM per photon with respect to the EC or the PC, $L_{y,j}^{(\rm int,EC/PC)}/N_j$, can be written by means of the generic formula
\begin{equation}\label{eq:Lyphoton}
\frac{L_{y,j}}{N_j}= \frac{\int\psi_j^* \hat L \psi_j dxdy dt'}{\int |\psi_j|^2dxdy dt'} ,
\end{equation}
with $\hat L = -k_j x$, $\hat L =iv_jt'\partial_x$ or $\hat L =i(v_jt'\partial_x- x\partial_{v_jt'})$, respectively,  and a similar expression with the differences of the integrals
at $z$ and 0 in the numerator and denominator for $\Delta L_{y,j}/\Delta N_{j}$, $\Delta L_{y,j}^{\rm (int,EC/PC)}/\Delta N_{j}$.

Equation (\ref{eq:Ly}) indicates that the total t-OAM of the distorted STOV vanishes. The numerical simulations of the SHG process show that its t-OAM continues to be zero, and that of the generated second harmonic is also zero (then not shown), hence the total t-OAM is conserved. The intrinsic t-OAM with respect to the EC, $L_{y,1}^{\rm(int,EC)}+ L_{y,2}^{\rm(int,EC)}$, and to the PC, $L_{y,1}^{\rm(int,PC)}+ L_{y,2}^{\rm(int,PC)}$, are both non-zero and turn out to be also conserved. As such, scaling laws of the form of Eq. (\ref{scaling2}) hold for both.

If the STOV were perfectly elliptical, the SHG process would not break this symmetry, and the harmonic field would also be an elliptical STOV. Elliptical STOVs have well-known values of their intrinsic t-OAM per photon $L_{y}^{\rm(int,EC)}/N=(\ell/2)\gamma$ \cite{porras2023transverse} for the EC choice, or $L_{y}^{\rm(int,PC)}/N=(\ell/2)(\gamma + 1/\gamma)$ for the PC choice \cite{Bliokh2023}, where $\gamma$ is the STOV ellipticity. Thus, since the intrinsic t-OAM of the driver does not change in the process, the general scaling law reduces to that  of intrinsic t-OAM per photon, $L_{y,2}^{\rm(int,EC/PC)}(z)/N_2(z) =2 L_{y,1}^{\rm(int,EC/PC)}(0)/N_1(0)$, and hence to $\ell_2=2 \ell_1$. In this ideal case, scaling TC is the same as intrinsic t-OAM conservation with respect to both the EC and the PC.

For a generic spatiotemporal driver, as the distorted STOV, the intrinsic t-OAM per photon changes in the process, as in Figs. \ref{fig:FIG2}(b) and (c) for the EC and PC choices (red and blue dashed curves for the driver and the harmonic), and are not given by the above formulas for elliptical STOVs. The  harmonic intrinsic t-OAM per photon at no $z$ is twice that of the driver at $z=0$, except with complete depletion, as indicated by the diagonal arrow. The solid red and blue curves in (b) and (c) are numerical verifications that $L_{y,2}^{\rm(int,EC)}(z)/N_2(z) = 2\Delta L_{y,1}^{\rm(int,EC)}(z)/\Delta N_{1}(z)$ and $L_{y,2}^{\rm(int,PC)}(z)/N_2(z) = 2\Delta L_{y,1}^{\rm(int,PC)}(z)/\Delta N_{1}(z)$, as indicated by the vertical arrow, as result of the conservation of the intrinsic t-OAM with respect to the PC and EC centroids.
$$\star\star\star$$

Thus, TC and OAM per photon are not physical magnitudes that allow us to infer OAM conservation with general spatiotemporal fields except very particular situations.
Direct, experimental measurement of the values of the OAM in Eq. (\ref{scaling2}) is very difficult, particularly the t-OAM has never been measured directly. Alternatively, OAM can be evaluated from the experimentally reconstructed fields in space and time, as we have done numerically. An important difference with the scaling $L_q^{\rm out}/N_q^{\rm out} = q L_1^{\rm in}/N_1^{\rm in}$ particular to LG pulses and other symmetry-preserving situations, is that the general scaling law in Eq. (\ref{scaling2}) involves not only the input driver but also the output driver.

\section{Scaling law with undepleted driver and extension to HHG}

The last observation posses the problem of evaluating $L_1^{\rm out}-L_1^{\rm in}$ and $N_1^{\rm out}-N_1^{\rm in}$ when depletion of the driver is barely observable, or not observable at all. Let us first consider an infinitesimal process of perturbative HG in which the driver experiences a change $\psi_1\rightarrow \psi_1 + d \psi_1$, and respective modifications of the OAM and the number of photons, $dL_1$ and $dN_1$. The scaling law then reads
\begin{equation}\label{undepleted}
\frac{L_q^{\rm out}}{N_q^{\rm out}} = q \frac{dL_1}{dN_1} ,  \quad {\rm (undepleted\,\,driver)}
\end{equation}
where the changes in $dL_1$ and $dN_1$ can be straightforwardly evaluated from Eqs. (\ref{eq:Lz}), or (\ref{eq:Ly}), and (\ref{eq:energy}), leading to the derivative of OAM with respect to the number of photons as
\begin{equation}\label{quotient}
\frac{dL_1}{dN_1} = \frac{{\rm Re}\int  d \psi_1^*\hat L  \psi_1 dxdy dt'}{{\rm Re}\int d \psi_1^* \psi_1  dxdy dt'}.
\end{equation}
The scaling law (\ref{undepleted}) with (\ref{quotient}) for OAM conservation may be easier to verify than Eq. (\ref{scaling2}), but requires, nevertheless, evaluation of the driver change $d\psi_1$ (a very small change compared to $\psi_1$). Again, the OAM per photon of the harmonic is not $q$ times that of the driver, but $dL_1/dN_1$ for OAM conservation: As seen in Figs. \ref{fig:FIG1}(b3) and \ref{fig:FIG2}(c), the dashed blue curve about $z=0$ is not twice the dashed red curve also about $z=0$ with OAM conservation.

This fact has recently been noticed in Ref. \cite{porras2025arxivPRL} for the intrinsic t-OAM in HHG in a gas-jet driven by a focused STOV, which amounts a tilted lobed field driver in the focal region, where the gas jet is placed. When it is placed at the focus, the high harmonics at the far-field are STOVs, all them with the same TC, hence with non-scaling TC. The intrinsic t-OAM per photon scales with harmonic order, but the scaled quantity is not the intrinsic t-OAM of the driver \cite{porras2025arxivPRL}, which is in line with Eq. (\ref{undepleted}).

\begin{figure*}[t]
    \centering
\includegraphics[width=1\linewidth]{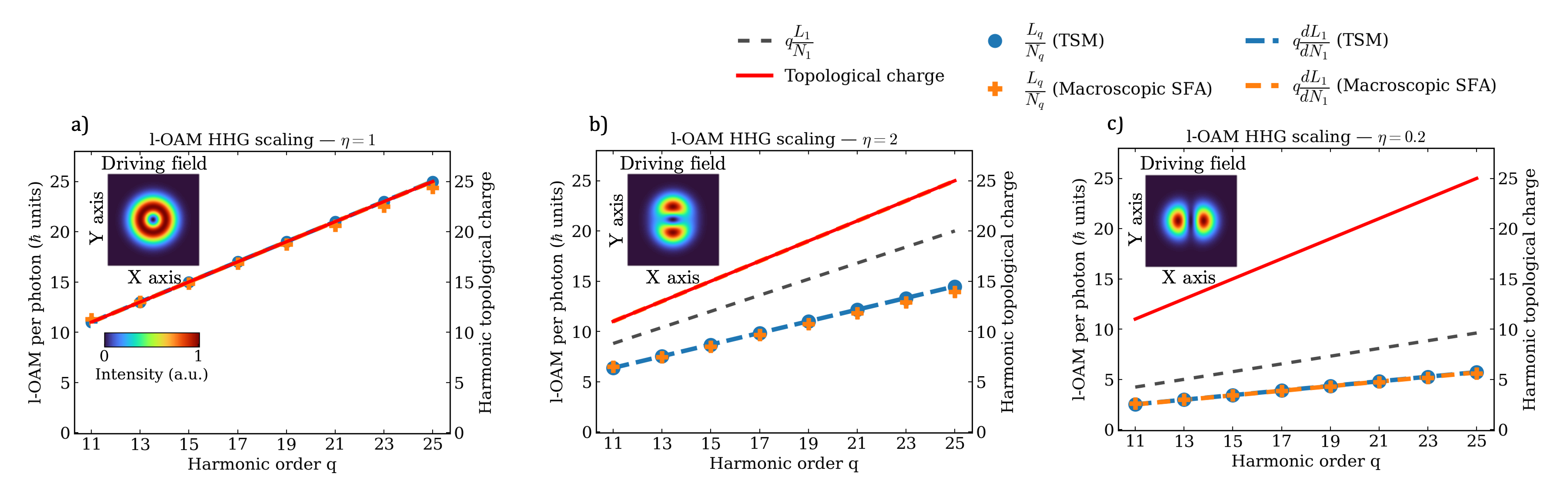}
    \caption{Harmonic l-OAM per photon scaling in HHG fitting the scaling law Eq. (\ref{undepleted2}), as evaluated from the TSM (blue) and macroscopic SFA approximation (orange). HHG is driven by a 800 nm wavelength pulsed longitudinal vortex $\psi_1(x,y,t') = A(\frac{x}{x_0} + i\eta \frac{y}{y_0})e^{-\frac{x^2}{x_0^2}-\frac{y^2}{y_0^2}-\frac{t'^2}{t_0^2}}$ of TC $\ell =1$, with $A$ such that the  peak field intensity is $1.4\times10^{14}\:\textrm{W/cm}^2$, $x_0=y_0=30\:\mu$m and $t_0=26.11$ fs. The vortex is deformed through the parameter $\eta$ as indicated. The intensity of the driving fields are shown in the insets. The driver TC and OAM per photon scaling are shown for comparison.}
    \label{fig:FIG3}
\end{figure*}

Indeed, in the same way as for HG, if the OAM is conserved along with the energy in a HHG process, we would have
\begin{equation}
    \frac{L_q^{\rm out}}{N_q^{\rm out}} = q \frac{L_1^{\rm out}- L_1^{\rm in} +\sum_{p\neq q} L_p^{\rm out}}{N_1^{\rm out}- N_1^{\rm in} +\sum_{p\neq q} N_p^{\rm out}},
\end{equation}
where $p$ is the order of all high-order harmonics except $q$.
Since driver depletion is very  small, we obtain, analogously to Eq. (\ref{quotient}),
\begin{equation}
 \frac{L_q^{\rm out}}{N_q^{\rm out}} = q \,\frac{{\rm Re}\int d\psi_1^* \hat L \psi_1 dxdydt' - \sum_{p\neq q}\int d\psi_p^* \hat L d\psi_p dxdydt'}{{\rm Re}\int d\psi_1^* \psi_1 dxdydt' - \sum_{p\neq q}\int d\psi_p^* d\psi_p dxdydt'}.
\end{equation}
The sums in the numerator and denominator are infinitesimal (very small) with respect to the first terms, and can be neglected. We then obtain approximately the same scaling law
\begin{equation}\label{undepleted2}
\frac{L_q}{N_q}\simeq  q \frac{dL_1}{dN_1}\quad {\rm (HHG)},
\end{equation}
with $dL_1/dN_1$ given by Eq. (\ref{quotient}), and where we have omitted "out" for the generated harmonics to lighten the notation. As in undepleted HG, conservation of OAM in HHG driven by longitudinal or transverse vortices does not require that the OAM per photon of a harmonic field is $q$ times that of the driver, but the right hand side of Eq. (\ref{quotient}). In \cite{porras2025arxivPRL}, $L_{y,q}^{\rm (int,EC/PC)}/N_q$ was evaluated, but $dL_{y,1}^{\rm (int,EC/PC)}/dN_1$ was not evaluated to check whether the scaling law (\ref{undepleted2}) holds, and hence whether the intrinsic t-OAM is conserved in the HHG process.

Here we prove analytically OAM conservation using the thin slab model (TSM) \cite{l1992calculations,hernandez2015quantum} of HHG in a gas jet with arbitrary driving fields by demonstrating Eq. (\ref{undepleted2}) within this model. The proof holds for the l-OAM, the t-OAM and its intrinsic part. OAM conservation is next supported by numerical simulations using the macroscopic strong field approximation (SFA) model \cite{lewenstein1994theory}.

First, we have evaluated approximately $dL_1/dN_1$ based on energy conservation. The input driver is assumed to undergo the transformation $\psi_1 =|\psi_1|e^{i\phi_1} \rightarrow \psi_1^{\rm(out)}=|\psi_1|e^{i\phi_1} + |d\psi_1|e^{i\phi_1}$ in which its phase $\phi_1$ is substantially unaltered. If the energy in any small area $dxdy$ of the thin gas jet is conserved, $|\psi_1|^2 =|\psi_1^{\rm (out)}|^2 + \sum_{q\neq 1}|\psi_q|^2$, or $|\psi_1^{\rm (out)}|\simeq |\psi_1|[1-(1/2)\sum_{q\neq 1}|\psi_q|^2/|\psi_1|^2]$ as the HHG efficiency is very small. The small change of the driver is then
\begin{equation}\label{driverchange}
d\psi_1\simeq |d\psi_1|e^{i\phi_1}\simeq- \frac{1}{2|\psi_1|}\sum_{q\neq 1}|\psi_q|^2 e^{i\phi_1} .
\end{equation}

In the TSM, the harmonic fields are approximated by $\psi_q \simeq A_q |\psi_1|^{q_{\rm eff}} e^{iq\phi_1}$, where $q_{\rm eff}$ is the nonperturbative scaling parameter (typically
$q_{\rm eff} \sim 3.5$ \cite{l1992calculations}), $A_q$ are harmonic amplitudes, and the intrinsic phase is neglected. This approximation leads to
\begin{equation}\label{driverchange2}
d\psi_1\simeq -\frac{1}{2}|\psi_1|^{2q_{\rm eff}-1}\sum_{q\neq 1}|A_q|^2 e^{i\phi_1}.
\end{equation}
The harmonic OAM per photon of the generated harmonics is then
\begin{equation}\label{uno}
    \frac{L_q}{N_q} \simeq \frac{\int |\psi_1|^{q_{\rm eff}}e^{-iq\phi_1}\hat L\,|\psi_1|^{q_{\rm eff}} e^{iq\phi_1}dxdydt'}{\int |\psi_1|^{2q_{\rm eff}} dxdydt'},
\end{equation}
and $q$ times the derivative of the driver OAM with  respect to the number of photons is given by
\begin{equation}\label{dos}
    q\,\frac{dL_1}{dN_1} \simeq q\frac{{\rm Re}\int |\psi_1|^{2q_{\rm eff}-1}e^{-i\phi_1}\hat L \,|\psi_1|e^{i\phi_1}dxdydt'}{\int|\psi_1|^{2q_{\rm eff}}dxdydt'},
\end{equation}
where $\hat L= -i (x\partial_y - y\partial_x)$ for the l-OAM, $\hat L=-k_q x$ for the t-OAM, $\hat L =ict'\partial_x$ for the intrinsic t-OAM about the EC, $\hat L = i(ct'\partial_x - x\partial_{ct'})$ for the intrinsic t-OAM about the PC, and we have assumed $n_i\simeq 1$. Equality of $L_q/N_q$ in Eq. (\ref{uno}) and $q\,dL_1/N_1$ in Eq. (\ref{dos}) is obvious for the total t-OAM since $k_q=qk_1$. For the l-OAM and the intrinsic t-OAM we use that $\hat L(fg) = (\hat L f)g + f(\hat L g)$ with $f= |\psi_1|^{q_{\rm eff}-n}e^{i(q-n)\phi_1}$ and $g=|\psi_1|e^{i\phi_1}$ starting with $n=1$ up to $q$ times, which results in
\begin{eqnarray}
    \frac{L_q}{N_q} &\simeq& \frac{\int|\psi_1|^{q_{\rm eff}+q}\hat L |\psi_1|^{q_{\rm eff}-q}dxdydt'}{\int|\psi_1|^{2q_{\rm eff}}dxdydt'} \nonumber \\
    &+& q \,\frac{ \int  |\psi_1|^{2q_{\rm eff}-1}e^{-i\phi_1}\hat L |\psi_1|e^{i\phi_1}dxdydt' }{\int|\psi_1|^{2q_{\rm eff}}dxdydt'}.
\end{eqnarray}
The first term is seen to vanish using $\hat L f^m = m f^{m-1}\hat L f$ and integrations by parts, the second term must be real as $L_q/N_q$ is the expectation value of Hermitian operators. Then, $L_q/N_q$ in Eq. (\ref{uno}) is equal to $qdL_1/N_1$ in Eq. (\ref{dos}) for arbitrary drivers within the TSM model, which amounts conservation of the l-OAM and the intrinsic t-OAM about the EC and PC in HHG as described by this model.

\begin{figure*}[t!]
    \centering
\includegraphics[width=1\linewidth]{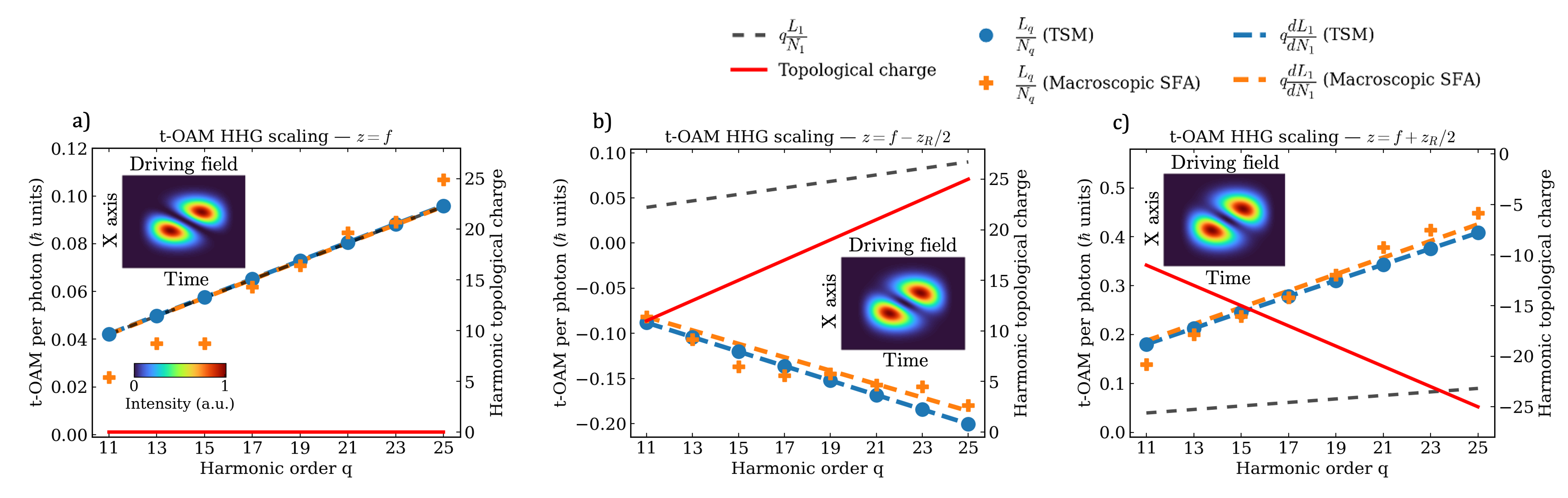}
    \caption{Harmonic intrinsic t-OAM per photon scaling in HHG fitting Eq. (\ref{undepleted2}), as evaluated from the TSM (blue) and macroscopic SFA approximation (orange). In (a) HHG is driven by the 800 nm  tilted-lobed field $\psi(x,y,t')= A (t'/t_0 - x/x_0)e^{-\frac{x^2}{x_0^2}-\frac{y^2}{y_0}-\frac{t'^2}{t_0^2}}e^{\frac{ik_1x^2}{2f}}$, with $x_0=y_0=30\:\mu$m and $t_0=26.11$ fs, and $A$ such that the peak intensity is $1.4\times10^{14}\:\textrm{W/cm}^2$, obtained by focusing a STOV of TC $\ell=1$ with a focal length $f=250$ mm. In (b) and (c) the driver is back- and forward- propagated from the focus distances $z_R/2=k_1x_0^2/4$. See \cite{Porras_nano_2025} for analytical expressions. These driving fields are the same as those used in \cite{porras2025arxivPRL}. The intensity of the driving fields are shown in the insets. The driver TC and OAM per photon scaling are shown for comparison.}
    \label{fig:FIG4}
\end{figure*}

We have corroborated OAM conservation by means of advanced numerical simulations of HHG driven by a number of relevant fields based on the SFA. We have considered HHG in an infinitesimally thin atomic hydrogen gas target, a valid approximation for short (compared to the Rayleigh length of the driving field) and low density targets. We calculate the near harmonic field $\psi_q$ by discretizing the target in a regular grid of individual emitters, where the HHG emission is obtained through the SFA without saddle point approximations \cite{perez2009harmonic}. Within this model we account for all trajectories ---i.e., short and long trajectories--- in the HHG process. From the macroscopic harmonic near field $\psi_q$, we obtain its OAM per photon $L_q/N_q$ with Eqs. (\ref{eq:Lzphoton}) or (\ref{eq:Lyphoton}) with the corresponding $\hat L$ operator, then we calculate the driver change $d\psi_1$ with Eq. (\ref{driverchange}) and, finally, $dL_1/dN_1$ with Eq. (\ref{quotient}). In all HHG numerical simulations we use a 800 nm wavelength driver of peak intensity $1.6\times10^{14}\:\textrm{W/cm}^2$ and a $\sin^2(\pi t/t_\textrm{max})$ temporal envelope, with $t_\textrm{max} = 32\:\textrm{optical cycles}$ (corresponding to $t_0\approx 26.11\:\textrm{fs}$ of a Gaussian envelope $e^{-t^2/t_0^2}$). The expressions of the drivers and their defining parameters are detailed in the captions of Figs. \ref{fig:FIG3} and \ref{fig:FIG4}. The results presented here are for an atomic hydrogen target, but they are easily extrapolated to any other noble gas.

In Fig. \ref{fig:FIG3} the driving fields are (a) a pulsed LG beam, (b) and (c) differently distorted pulsed LG beams, all with $\ell=1$ and l-OAM. The values of $qdL_1/dN_1$ (orange dashed line) and $L_q/N_q$ (orange crosses) computed from the macroscopic SFA coincide in (a), (b) and (c), showing l-OAM conservation, and coincide with the prediction that $qdL_1/N_1$ (blue dashed line) aquals to $L_q/N_q$ (circles) for l-OAM conservation from the TSM. Only for the perfect LG driver in (a), $qdL_1/dN_1$ equals to the scaling $qL_1/N_1$ of the driver OAM per photon (grey dashed line) and the scaling of the TC (red line). For arbitrary longitudinal vortices, driver OAM per photon and TC scaling are not related to l-OAM conservation.
%For arbitrary disturbed LG driving beams, Figs. \ref{fig:FIG3} b) and c), l-OAM conservation in the HHG process is only compatible with the scaling of $dL_1/dN_1$ and unintelligible from the l-OAM per photon scaling of the driving field $qL_1/N_1$ nor the topological charge scaling of the harmonic field.

In Fig. \ref{fig:FIG4} we elaborate a similar analysis for the intrinsic t-OAM with respect to the EC when the driving fields are tilted lobed fields, obtained by focusing an elliptical STOV with $\ell=1$ when the gas-jet is placed (a) at focus, (b) before focus, and (c) after focus. These driving fields are the same as in \cite{porras2025arxivPRL} to elucidate OAM conservation. In all three cases the good fitting of $qdL_1/dN_1$ to $L_q/N_q$ evaluated from the advanced numerical simulations (orange dashed line and crosses) is compatible with intrinsic t-OAM conservation. With these drivers, the conservation predicted analytically by the TSM (blue dashed lines and circles) differs slightly from that of the SFA. As with perfect LG drivers, when target is located at the focus in (a), the driver presents a symmetry such that its intrinsic t-OAM per photon is preserved in the HHG process. Then, the scaling law for OAM conservation reduces to $L_q/N_q=qL_1/N_1$ (gray dashed line). However, contrary to the LG driver, the TC of all near-field harmonics is the same (red line), developing into far-field harmonic STOVs all them of unit TC, and allowing for the generation of attosecond STOVs \cite{porras2025arxivPRL}. Out of focus, as in (b) and (c), scaling of the driver t-OAM per photon (gray dashed lines) and of the TC (red lines) differ from that of $dL_1/dN_1$, but this is alien to OAM conservation. This is dramatically illustrated by scaling of the same or opposite slopes of $L_1/N_1$ and $dL_1/N_1$ before and after the focus in (b) and (c), and by scaling of TC an $dL_1/N_1$ of opposite slopes in (b) and (c).
%We note that Figs. \ref{fig:FIG3} and \ref{fig:FIG4} reveal that the harmonic topological charge scaling law is not indicative of l-OAM or t-OAM conservation in general. In disturbed LG-driven HHG [Figs. \ref{fig:FIG3} b) and c)], the topological charge scaling of the harmonics generated differs from the harmonic l-OAM scaling $L_q/N_q$. Both scalings only coincide if the driving LG beam is perfectly symmetric [Fig. \ref{fig:FIG3} a)]. In focused STOV-driven HHG the difference between harmonic topological charge scaling and t-OAM conservation is further emphasized. When HHG is driven with the target located at focus [Fig. \ref{fig:FIG4} a)], the topological charge (red line) of all the emitted harmonics is the same, developing into a unit topological charge in the far field \cite{porras2025arxivPRL}. If the target is located before or after focus, the difference between topological charge and t-OAM scaling becomes more dramatic as they display opposite sign behaviours, positive topological charge scaling  and negative t-OAM slope in HHG driven before focus [Fig. \ref{fig:FIG4} c)], and vice versa in HHG driven after focus [Fig. \ref{fig:FIG4} c)]. Therefore, we can conclude that harmonic topological charge scaling is not a good indicative of OAM conservation in non-linear frequency up-conversion processes.
The examples in Figs. \ref{fig:FIG3} and \ref{fig:FIG4} evidence, in short, the variety of relations between TC, driver OAM per photon, and harmonic OAM scaling, all them compatible with OAM conservation, and unintelligible from the rigid scaling law for LG drivers.
%Only with a perfect LG driver, TC, driver OAM and harmonic OAM per photon fitting $\Delta L/\Delta N$ coincide [Fig. \ref{fig:FIG3}(a)]. The situation is similar with the focused STOV at the focal plane, but the TC is zero for the driver and all harmonics at the focus [Fig. \ref{fig:FIG3}(a)], though they all develop unit TC at the far field \cite{porras2025arxivPRL}. With distorted LG beams [Fig. \ref{fig:FIG3}(b) and (c)], TC and OAM per photon scaling do not inform about OAM conservation. The same for focused STOV out of focus [Fig. \ref{fig:FIG4}(b) and (c)], where TC and OAM per photon can even have opposite signs.

\section{Conclusion}

To conclude, we believe it has been timely to critically review the identification of TC or OAM per photon scaling with harmonic order with OAM conservation in HG and HHG processes. With a generic spatiotemporal field, neither TC nor OAM per photon, scaling or not, are connected to OAM conservation. TC scaling occurs in most cases, but not all. TC scaling is a topological transformation dictated by typical laws of harmonic generation. OAM per photon scaling occurs only in processes in which this magnitude remains invariable, as for perfect LG beams or perfect STOVs (if diffraction effects in the latter are negligible). In this sense, LG beams and STOVs in experiments are never perfect (as in the examples in this paper), even present strong distortions. Then, a genuine demonstration of the conservation of in a specific platform of HG or HHG would require the verification that the converted OAM per converted number of photons scales with the harmonic order, which requires not only information about the input driver, but also the output driver. Our results establish a basis for a correct interpretation of present and future, theoretical or experimental, research on nonlinear processes driven by structured light carrying OAM.

\bibliography{bibliography}

\end{document}